\documentclass[prl,aps,twocolumn,showpacs,amsmath,amssymb,superscriptaddress]{revtex4}

\usepackage[dvips]{graphicx}
\usepackage{dcolumn}
\usepackage{bm}

\begin{document}

\preprint{}

\title{Universality Class of Fiber Bundle Model on Complex Networks}

\author{Dong-Hee Kim}
\affiliation{Department of Physics, Korea Advanced Institute of Science and Technology, Daejeon 305-701, Korea}
\author{Beom Jun Kim}
\affiliation{Department of Molecular Science and Technology, Ajou University, Suwon 442-749, Korea}
\author{Hawoong Jeong}
\affiliation{Department of Physics, Korea Advanced Institute of Science and Technology, Daejeon 305-701, Korea}

\begin{abstract}

We investigate the failure characteristics of complex networks 
within the framework of the fiber bundle model subject to the 
local load sharing rule in which the load of the broken fiber
is transferred only to its neighbor fibers.   
Although the load sharing is strictly local, it is found that
the critical behavior belongs to the universality class of the global 
load sharing where the load is transferred equally to all fibers in the
system. From the numerical simulations and the analytical 
approach applied to the microscopic behavior, it is revealed that
the emergence of a single dominant hub cluster of broken fibers causes 
the global load sharing effect in the failure process. 

\end{abstract}

\pacs{62.20.Mk,89.75.Hc,05.70.Jk,64.60.Fr}

\maketitle

The fiber bundle model (FBM) has been studied for many years
in order to explain a variety of failure phenomena 
caused by cascades~\cite{review}. 
In the FBM, composed of $N$ heterogeneous fibers put on a lattice,
a fiber at the $v$th site is broken if the load $\sigma_v$ is larger than 
the threshold value $\sigma^{th}_v$ assigned to the fiber 
following a given probability distribution function.
When the fiber is broken the load which was supported by the broken 
fiber is shared among intact fibers following a load sharing rule.
The two most frequently studied rules are 
global load sharing (GLS)~\cite{GLS}, 
in which the load of a broken fiber is equally shared 
with all intact fibers in the whole system, and local 
load sharing (LLS)~\cite{LLS}, which allows only the nearest intact fibers 
to carry the load of a broken fiber.

Depending on which load sharing rule is used, the FBM has been
shown to exhibit totally different behaviors: For GLS, the FBM 
has been found to have a phase transition toward the global failure 
as the external load per fiber  ($\bar\sigma$)
is increased beyond the nonzero critical point 
$\bar\sigma_c$~\cite{Zapperi}. 
It has been known that the avalanche size distribution for the GLS
follows the power-law form with the universal exponent 
$-5/2$~\cite{Hemmer1,Kloster1}, and the universality class 
has been identified from the measurement of critical 
exponents~\cite{critical}.  
In contrast, the FBM under the LLS rule has been shown to
belong to a completely different universality class; i.e., 
the critical value of the load approaches zero as $N$
is increased following the form  $\bar\sigma_c \sim 1/\ln(N)$
~\cite{Smith-Gomez,Hansen1}. Recently, the transition 
between the GLS and LLS regime has been studied 
using modified load sharing rules~\cite{crossover}.

Until very recently, most studies of the FBM have been performed
on regular lattice structures. In a general perspective
beyond the fracture of material, however, cascading failure 
triggered by overloading happens also in real-world network 
systems~\cite{fbmnetwork,Holme1}. 
For instance, the recent blackout in the United States and 
Canada was caused by cascading breakdown of elements through a power grid.  
The scenario of major blackout is very similar to the idea of the 
FBM; once one element fails, then the neighbor of the elements
fail by the increased load from the failure. 

In this Letter, we numerically and analytically study the FBM subject to the
LLS rule on various network structures: i.e.,  
the Erd\"os-R\'enyi (ER) model of a random network~\cite{ER}, 
the Watts-Strogatz (WS) model of a small-world network~\cite{WS}, 
and the static model of a scale-free network~\cite{static_model}. 
Even though the model studied in this work obeys strictly the LLS rule, 
it is found that the FBM on complex networks exhibits
completely different universality:
the critical behavior, the avalanche size distribution, and 
the form of failure probability function coincide with 
those of the FBM under the GLS rule.

First, we briefly describe the FBM under the LLS rule 
on complex networks. A version of blackout scenarios for cascading 
breakdown of power plants from overloading is very intuitive to understand 
how the FBM on complex networks works. Fibers attached to the vertices 
of underlying networks act as power plants, and the external 
load on fibers can be regarded as the demand for electric power.
If the demand for electric power exceeds the capacity of a power plant, 
the power plant gets disconnected from the network and the demand is 
transferred to neighboring power plants through the transmission lines, 
the edges of the network. In this model, the underlying network is 
rigid while the fibers attached to vertices are damaged. 

\begin{figure}
\centering{\resizebox*{0.5\textwidth}{!}{\includegraphics{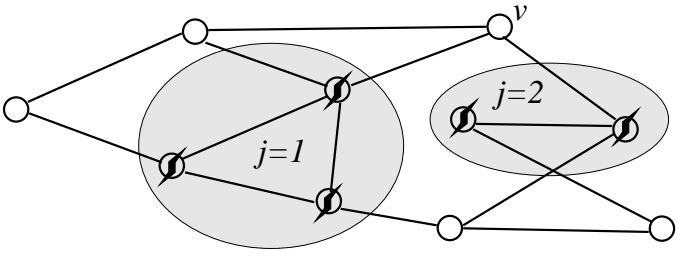}}}
\caption{
\label{fig:lls} 
The LLS rule applied for the FBM on a complex
network. The vertices bound to broken fibers and intact fibers are 
denoted by the broken symbols and the empty circles, respectively. }
\end{figure}

The local load transfer of broken fibers through the edges of 
the underlying network is governed by the LLS rule.
Under a non-zero external load $N\bar\sigma$, the actual load $\sigma_v$
of the intact fiber $v$ is given by the sum of 
$\bar\sigma$ and the transferred load from neighboring broken fibers.
To systematically handle the local load transfer from broken fibers 
to intact fibers, we define the load concentration factor 
$K_v \equiv \sigma_v/\bar\sigma$ as
$K_v = 1 + {\sum_j}^\prime m_j / k_j$,
where the primed summation is over the cluster of broken fibers
directly connected to $v$, $m_j$ is the number of broken fibers 
in the cluster $j$, and $k_j$ is the number of intact fibers 
directly connected to $j$. 
A simple example is shown in Fig.~\ref{fig:lls}, which contains two 
clusters of broken fibers, $m_1 =3$, $k_1 = 4$, and $m_2 = 2$, $k_2 = 3$. 
If $\sigma^{th}_v < (1+3/4+2/3)\bar\sigma$, the fiber at $v$ will be broken 
and join the clusters of broken fibers. 
We note that this is the generalization of the 1D model 
($k_j=2$)~\cite{Newman1,Zhang1}.

To explore all values of the external load $N\bar\sigma$, we increase
$\bar\sigma$ quasistatically starting from zero. 
In a finite-sized system, the infinitesimal increment $\delta$ of 
$\bar\sigma$ is expressed as
$\delta = \min_v \left[ \frac{\sigma^{th}_v}{K_v} - \bar\sigma \right]$
with the minimization ($\min_v$) for all intact fibers. 
This condition is equivalent to the increase of $\bar\sigma$
just enough to break only the weakest intact fiber, which is the minimal
condition to trigger an avalanche.

Our numerical scheme is as follows: 
The threshold value of the load $\sigma_v^{th} \in [0,1]$ is assigned to
each fiber following the uniform distribution function~\cite{comment}. 
Start from $\bar\sigma = 0$ and repeat the following two steps: 
(i) Increase $\bar\sigma$ by the infinitesimal increment $\delta$. 
(ii) Following the LLS rule,
break the fibers with $\sigma_v^{th} < K_v\bar\sigma$ iteratively until 
no more fibers break. 
For each increment of $\bar\sigma$, the size $s(\bar\sigma)$ of 
the avalanche is defined as the number of broken fibers triggered 
by the increment. 
The surviving fraction $x(\bar\sigma)$ of fibers is 
the ratio of the number of remaining intact fibers to $N$ when 
the external load reaches $N\bar\sigma$, and thus is written as
\begin{equation}
\label{eq:x}
x(\bar\sigma)= 1-\frac{1}{N}\sum_{\sigma < \bar\sigma} s(\sigma). 
\end{equation}
We also measure the response function $\chi$, or the generalized 
susceptibility, by using 
\begin{equation}
\chi(\bar\sigma) = \left| \frac{dx}{d\bar\sigma} \right| 
= \frac{1}{N \Delta}\sum_{\bar\sigma < \sigma < \bar\sigma+\Delta} s(\sigma),
\end{equation}
where we choose the small enough value 
$\Delta = 0.0005$ for the numerical differentiation.
The critical value $\bar\sigma_c$ of the external load, which
is one of the key quantities of interest, is defined
from  the condition  of the global breakdown $x(\bar\sigma_c)=0$ .

\begin{figure}
\centering{\resizebox*{0.5\textwidth}{!}{\includegraphics{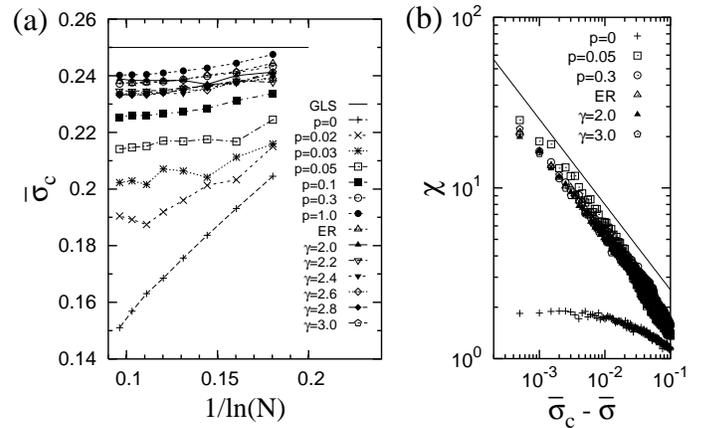}}}
\caption{
\label{fig:sigmac} 
(a) The system size ($N$) dependence of critical points ($\bar\sigma_c$)
for various networks with $N=2^8,2^9,\ldots,2^{15}$ vertices. 
(b) The susceptibility for the networks with $N=2^{14}$.
$p$ and $\gamma$ are the rewiring probability 
in the WS networks and the exponent of degree distribution 
$P(k) \sim k^{-\gamma}$ in the static model~\cite{static_model},
respectively. The data points are obtained 
from the averages over $10^4$ ($10^3$ for $N=2^{15}$) ensembles. }
\end{figure}

Figure~\ref{fig:sigmac} displays the critical value $\bar\sigma_c$
and the susceptibility $\chi$ 
for the FBM under the LLS rule on various network structures,
such as the local regular network, the WS network, the ER network,
and the scale-free networks.
Strikingly, we find that the critical behavior of the FBM on complex networks 
is completely different from that on a regular lattice. More specifically,
while $\bar\sigma_c$ for the FBM on a local regular network vanishes
in the thermodynamic limit and is described by $\bar\sigma_c \sim 1/\ln(N)$ 
for finite-sized systems (see the curve for $p=0$ in
Fig.~\ref{fig:sigmac}, corresponding to the WS network with the
rewiring probability $p=0$), $\bar\sigma_c$ for {\em all} networks except
for the local regular one does not diminish but converges to
a nonzero value as $N$ is increased.
Moreover, the susceptibility diverges at the critical point 
following $\chi \sim (\bar\sigma_c - \bar\sigma)^{-0.5}$, regardless of 
the networks, which is again in a sharp contrast to the local regular
network [see Fig.~\ref{fig:sigmac}(b)]. 
The critical exponent $0.5$ clearly indicates that the FBM under
the LLS rule on complex networks belongs to the same universality
class as that of the GLS regime~\cite{critical} 
although the load-sharing rule is strictly local.

\begin{figure}
\centering{\resizebox*{0.5\textwidth}{!}{\includegraphics{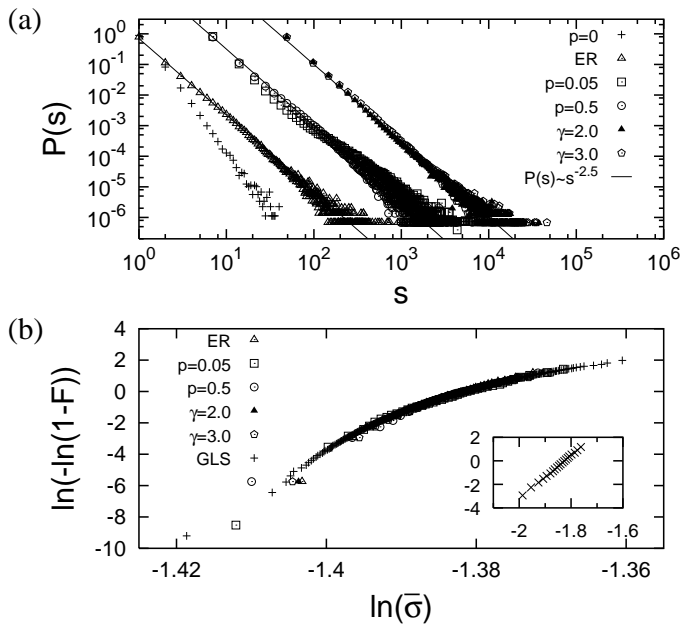}}}
\caption{
\label{fig:aval} 
(a) The avalanche size distribution $P(s)$ and (b) the  
test of failure probability $F$ for the Weibull form. 
The solid line in (a) indicate the result for the GLS regime having 
exponent $-5/2$. The inset in (b) shows the result for the LLS regime on
the regular network ($p=0$), which has clearly different form from others.
}
\end{figure}

The evidences that the LLS model on complex networks belongs to 
the universality class of the GLS model is also found in the avalanche size 
distribution $P(s)$: Unanimously observed power-law behavior
$P(s) \sim s^{-5/2}$ in Fig.~\ref{fig:aval} (a) for all networks
except for the local regular one (the WS network with $p=0$)
is in perfect agreement with the same behavior for the GLS 
case~\cite{Hemmer1}. On the other hand, the LLS model  for a regular 
lattice has been shown to exhibit completely different avalanche size
distribution~\cite{Hansen1,Kloster1}.

The failure probability $F(\bar\sigma)$ is defined as the probability that 
the whole system is broken when an external load 
$\bar\sigma$, or less, is applied.
In the LLS regime, the failure probability was studied to test the 
weak-link hypothesis~\cite{Harlow1,Leath1,Zhang1}. 
In the test for the Weibull form, one can see in Fig.~\ref{fig:aval} (b) 
that $F(\bar\sigma)$'s for complex networks fall on a common line which 
coincides with $F(\bar\sigma)$  for the well-known GLS case,  which is
very much different for LLS on regular lattices~\cite{Harlow1,Leath1,Zhang1} 
[see the inset in Fig.~\ref{fig:aval}(b)]. 
Consequently, we again confirm that the LLS model on complex
networks belongs to the same universality class as that of GLS.

\begin{figure}
\centering{\resizebox*{0.5\textwidth}{!}{\includegraphics{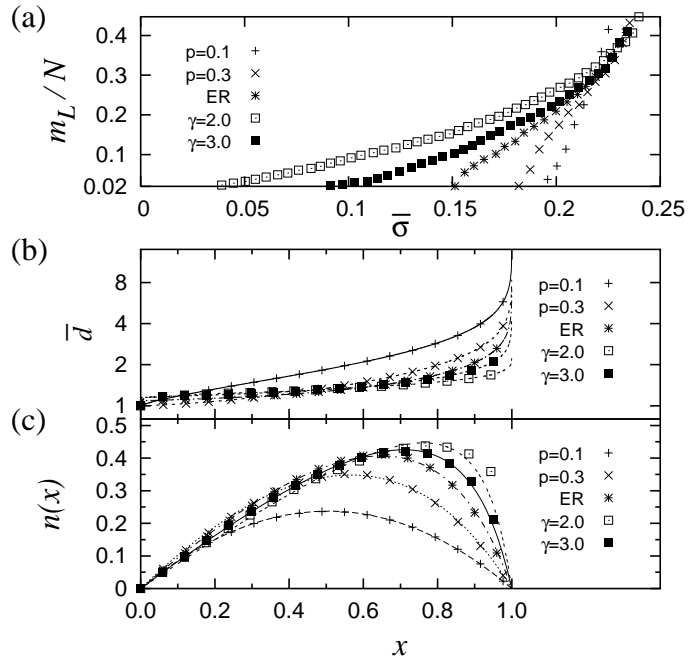}}}
\caption{ 
\label{fig:cluster}
(a) The largest cluster size $m_L$ in the failure process versus the external 
load $\bar\sigma$. 
(b) The average distance $\bar{d}$ from the 
DLC versus the surviving fraction $x$. 
(c) The fraction $n(x)$ of nearest intact fibers of the DLC 
as a function of the surviving fraction $x$.  
The symbols and the lines represent the numerical data and their fits to 
the curve of Eq.~(\ref{eq:nt}), respectively.
The data points are obtained for 
the networks with $N=2^{14}$ and averaged over $10^4$ ensembles. 
}
\end{figure}

What makes the LLS model on a complex network have identical 
critical behaviors of the GLS model?
In order to answer this question, it is helpful to investigate  
the microscopic details of the failure process. 
As the external load increases, the clusters of broken vertices 
form, grow, and merge into larger clusters, ultimately 
resulting in the emergence of a dominantly large cluster (we call it
DLC henceforth). 
In case of a regular lattice, all small clusters of 
broken fibers are roughly in an equal condition 
because of the underlying regular (and thus spatially uniform) topology. 
As the external load increases, all small clusters grow 
at roughly the same rate, and thus the DLC emerges abruptly.
In contrast, the emergence of a DLC on complex networks
is a gradual process because the DLC is formed at an early stage 
of loading and keeps growing continuously, as shown clearly 
in Fig.~\ref{fig:cluster}(a).
The above observation indicates that the growth of 
the DLC plays a dominant role in the failure of fibers on complex networks. 

Together with the early emergence and the gradual growth of the DLC, 
the small-world behavior~\cite{WS} in complex networks
provides a reasonable qualitative explanation of the GLS-like
behavior of the LLS model on complex networks.
More precisely, the small-world effect causes the average 
distance from the DLC to intact fibers to decay fast towards the value 
below two as the size of the DLC increases 
(or the surviving fraction $x$ decreases)
[see Fig.~\ref{fig:cluster}(b)], which implies that 
most intact fibers are very closely located to the DLC
and thus the system behaves similarly to the GLS model with
all fibers are separated by the unit distance to each other.

Finally, we apply the mean-field theory to the growth of the DLC.
Let us assume the situation that the system is
at the $t$th step of the load transfer from the DLC.
The surviving fraction is written as $x^{(t)}$,
and thus the cluster is composed of $N(1-x^{(t)})$ broken vertices
(we assume that there exists only one cluster) and has 
$k^{(t)}$ nearest neighbor intact vertices.
Following the LLS rule, the load $\sigma^{(t)}$ of the nearest intact vertices 
of the cluster satisfies the load conservation condition, which
yields
\begin{equation}
\label{eq:sigmat}
\sigma^{(t)} = \frac{1+n^{(t)}-x^{(t)}}{n^{(t)}}\bar\sigma ,
\end{equation}
where $n^{(t)} \equiv k^{(t)}/N$. 
Assuming that the load threshold $\sigma^{th}$ of a fiber 
is randomly distributed to the whole system following the cumulative 
distribution $P_{cum}(\sigma^{th})$, we establish the recursion 
equation for $x$, 
\begin{equation}
\label{eq:xt}
x^{(t+1)} = x^{(t)} - 
n^{(t)}\left[ 1- \frac{1-P_{cum}(\sigma^{(t)})}{x^{(t)}} \right]. 
\end{equation}
In order to solve the equations, we have to know the functional 
form of $n^{(t)}$, which is difficult to determine analytically.
Instead, we assume the simple rational functional form 
with two fitting parameters $a,b \in [0,1]$.
\begin{equation}
\label{eq:nt}
n^{(t)} \simeq bx^{(t)}(1-x^{(t)})/(1-ax^{(t)}),
\end{equation}
which is motivated as a generalization of the GLS form 
(corresponding to $a=b=1$).
The numerical data of $n^{(t)}$ fit very well to
the form~(\ref{eq:nt}) as shown in Fig.~\ref{fig:cluster}(c).

When the failure stops propagating, the fixed point
of the dynamics described by Eq.~(\ref{eq:xt}) satisfies
$x^{(t+1)}=x^{(t)} \equiv x^*$.
The uniform threshold distribution $P_{cum}(\sigma)=\sigma$,
together with Eqs.~(\ref{eq:sigmat})-(\ref{eq:nt}), results in 
\begin{equation}
\label{eq:xstar}
\bar\sigma(x^*) = \frac{bx^*(1-x^*)}{1+(b-a)x^*}.
\end{equation}
The critical value $\bar\sigma_c$ of the external load 
and the surviving fraction at $\bar\sigma_c$ are  then easily
obtained from the maximum of $\bar\sigma(x^*)$, yielding
\begin{equation}
\bar\sigma_c = b{x^*_c}^2, ~~
x^*_c = \frac{-1 + \sqrt{1+b-a}}{b-a} .
\end{equation}
Near the GLS regime, $b=1$ and $a=1-\epsilon$, one obtains
\begin{equation}
\bar\sigma_c \simeq 1/4 - \epsilon/8 ,~~~ x^*_c \simeq 1/2-\epsilon/8, 
\end{equation}
which indicates that the solution for the GLS regime is revisited 
at $\epsilon =0$ in a good agreement with the numerical simulation where
$\bar\sigma_c$ has always been found to be smaller than the GLS
value $\bar\sigma_c = 1/4$ (see Fig.~\ref{fig:sigmac}).  
In addition, combining Eqs.~(\ref{eq:sigmat}) and~(\ref{eq:xstar}), 
we obtain the critical behavior of $x^*$ near $\bar\sigma_c$, 
\begin{equation}
\chi = \left| \frac{dx^*}{d\bar\sigma} \right| \simeq (\bar\sigma_c - \bar\sigma)^{-1/2}, 
\end{equation}
confirming the universal exponent $-0.5$ for the GLS regime. 

In conclusion, we have investigated the FBM
on various complex networks including the WS network, the ER random network, 
and the scale-free network.
From numerical simulations and analytical approach, 
it has been found that although the LLS rule is strictly
applied, the critical behavior of failure exhibits the characteristics 
of GLS: The critical value of external load
is finite in the thermodynamic limit, the divergence of the susceptibility 
is described by the same critical exponent as in GLS; 
the avalanche size distribution and the statistics of 
failure display the unique behavior of the GLS regime. 

\acknowledgments
This work was supported by 
KOSEF-ABRL program (R14-2002-059-01002-0,DHK) 
and by Korea Research Foundation through 
Grant No. 2003-041-C00137 (BJK) and by KOSEF through
Grant No. R08-2003-000-10285-0 (HJ).


\begin{thebibliography}{}
\bibitem{comment} We also check that our main results does not 
change if we use the Weibull distribution instead of the uniform one.
\end{thebibliography}
\end{document}